\documentclass[journal]{IEEEtran}
\usepackage{float}
\usepackage{subfigure}
\usepackage{cite}

\ifCLASSINFOpdf
  \usepackage[pdftex]{graphicx}
\else
\fi
\begin{document}
%
\title{DARWIN: A Highly Flexible Platform for Imaging Research in Radiology}%
%
%

\author{Lufan Chang, Wenjing Zhuang, Richeng Wu, Sai Feng, Hao Liu, Jing Yu, Jia Ding, Ziteng Wang, Jiaqi Zhang
\thanks{Authors are with the Yizhun Medical AI Co.Ltd, Beijing, China}
\thanks{Emails: \{lufan.chang, wenjing.zhuang, richeng.wu, sai.feng, hao.liu, jing.yu, jia.ding, ziteng.wang, jiaqi.zhang\}@yizhun-ai.com}
\thanks{Manuscript received April 19, 2005; revised August 26, 2015.}
}

\maketitle

\begin{abstract}
To conduct a radiomics or deep learning research experiment, the radiologists or physicians need to grasp the needed programming skills, which, however, could be frustrating and costly when they have limited coding experience. In this paper, we present DARWIN, a flexible research platform with a graphical user interface for medical imaging research. Our platform is consists of a radiomics module and a deep learning module. The radiomics module can extract more than 1000 dimension features(first-, second-, and higher-order) and provided many draggable supervised and unsupervised machine learning models. Our deep learning module integrates state of the art architectures of classification, detection, and segmentation tasks. It allows users to manually select hyperparameters, or choose an algorithm to automatically search for the best ones. DARWIN also offers the possibility for users to define a custom pipeline for their experiment. These flexibilities enable radiologists to carry out various experiments easily.
\end{abstract}


%
\IEEEpeerreviewmaketitle

\section{Introduction}
%
%
%
%
\IEEEPARstart{M}{achine} learning becomes ubiquitous in recent medical imaging analysis research. The radiomics and deep learning techniques have facilitated numerous disease diagnosis with high precision \cite{mckinney2020international,review, wibmer2015haralick}.
There has been a series of applications of machine learning to medical imaging. Physicians can use machine learning techniques to explore the relationships between images processed using radiomics, clinical outcomes and radiation dose data to help improve cancer treatment with radiotherapy \cite{review}. Bayesian Network (BN), decision tree (DT), and Support Vector Machine (SVM) have been employed in various studies to predict 2-year survival \cite{Rapid, radiomicML, Predictout}. Numerous machine learning models and feature selection techniques for Non-small-cell Lung Carcinoma (NSCLC) cancer patients has been investigated in \cite{MLquant}. Machine learning models also facilitate many tasks such as lung nodules classification \cite{shen2015multi,hussein2017risk}, lesions detection \cite{yan20183d}, segmentation \cite{kamnitsas2017efficient, christ2016automatic}, and image registration \cite{balakrishnan2018unsupervised}.



Many machine learning platforms for general tasks have been developed by some high technology companies such as Google, Amazon, and Microsoft. These platforms greatly reduce duplication of efforts for developers. However, coding experience is needed for conducting experiments on these platforms, which makes it difficult for physicians with limited programming skills. Furthermore, these platforms do not focus on medical images, some requirements from physicians or radiologists may not be covered. 

A few platforms targeting medical images have come out to facilitate radiologists to carry out their research \cite{cid2017quantimage, zhang2015ibex, nioche2018lifex, yuan2019radiomics}. 
Imaging Biomarker Explorer (IBEX) \cite{zhang2015ibex} is a graphical-user-interface (GUI) based platform for medical image analysis. It integrates ROI labeler, radiomic feature extractor, and feature/model viewer. IBEX offers the flexibility for users to choose features, however, it only works on the Windows system, and the dependency files make the installation rather difficult. RayPlus \cite{yuan2019radiomics} has been proposed to solve this problem. RayPlus is a web-based platform across all systems. The features provided include five image filters with histogram, volumetric, morphologic, and texture features. These features could be  downloaded and further analyzed in R or SPSS.

These GUI based platforms enable users to do image analysis with minimal coding experience, however, still have notable weaknesses: 
1) Only certain types of images such as CT, PET are supported, which cannot satisfy the needs of physicians to study various types of images. 
2) Limited numbers of features are included, making it hard to generalize well on different tasks. 
3) Only radiomic features provided, and users still need to write programs on machine learning models in a way they can make clinical predictions using these features.
4) State of the art deep learning methods are not covered in these platforms.

Therefore we developed DARWIN: an artificial intelligence research platform for medical imaging. The platform integrates numerous data preprocessing methods, machine learning models, and data visualization methods with a web-based GUI. The platform does not have steep learning curves for radiologists to master it. 
The main contributions of the platform mainly lie on the following four aspects: 
1) incorporated PyRadiomics library\cite{pyradiomics} to extract abundant radiomics features. 
2)provided several deep learning models with hyperparameter tuning algorithms 
3) provided customized pipelines for different applications. 
4) provided a user-friendly interaction mode.

\begin{figure*}[htbp]
\centerline{\includegraphics[width=0.9\linewidth]{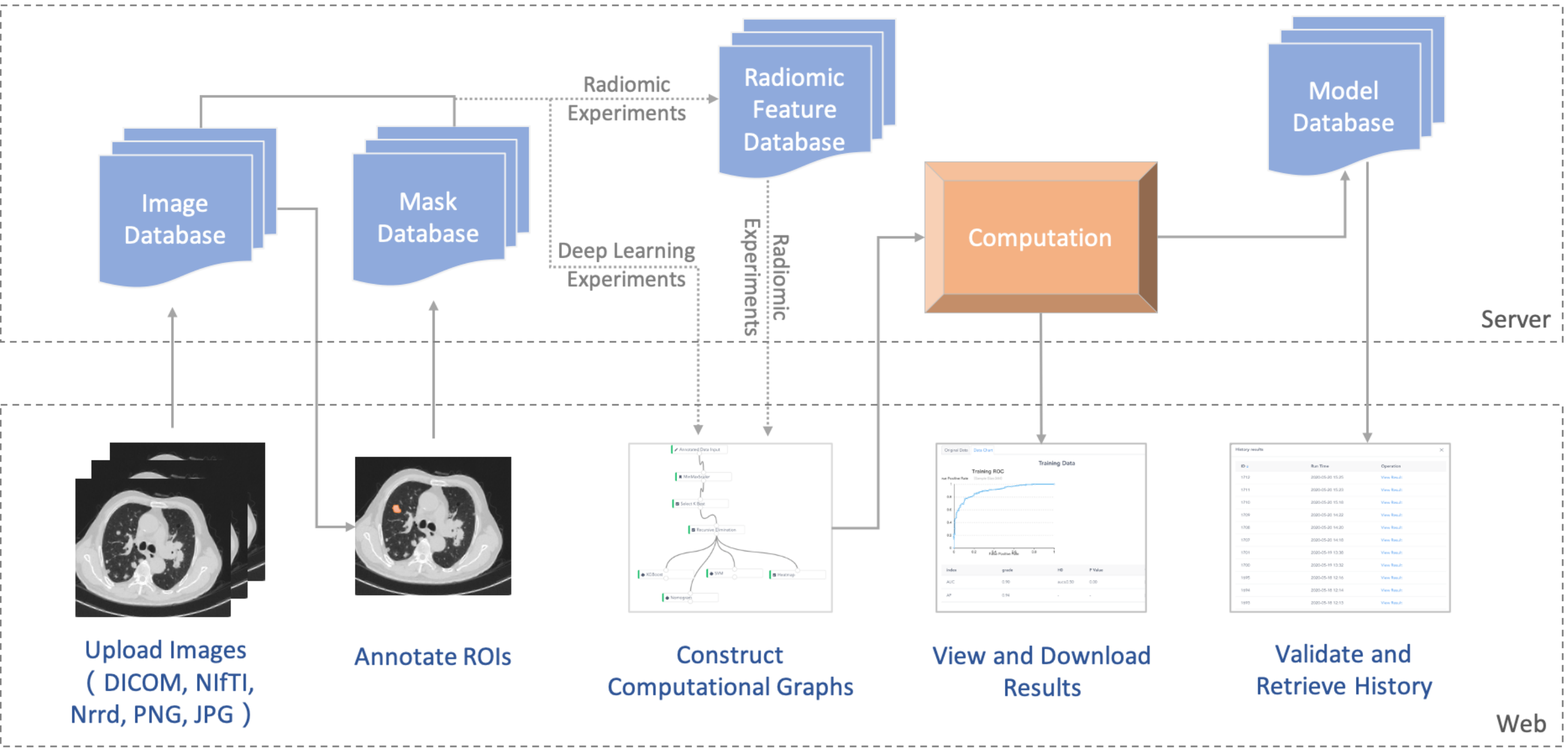}}
\caption{Computational Architecture of the DARWIN Platform. Users can use the web browser to upload the experiment images, annotate region of interests (ROIs), construct a computational graph, download results, and retrieve history records. The four databases at the server-side store the images, masks, radiomic features ,and models respectively. 
}
\label{computational_architecture}
\end{figure*}

\section{Overall Features of the Platform}
DARWIN research platform is a software mainly designed for radiologists to conduct machine learning research. It is convenient for physicians or radiologists to build an experiment pipeline to validate their hypotheses immediately. Fig \ref{computational_architecture} presents the architecture of DARWIN. It is based on Browser/Server (B/S) structure. Users can access the platform with a web browser on any operating system (Mac OSX, Windows, and Linux). It is implemented using Javascript, Python, and C++.

\begin{figure*}[htbp]
\centerline{\includegraphics[width=\linewidth]{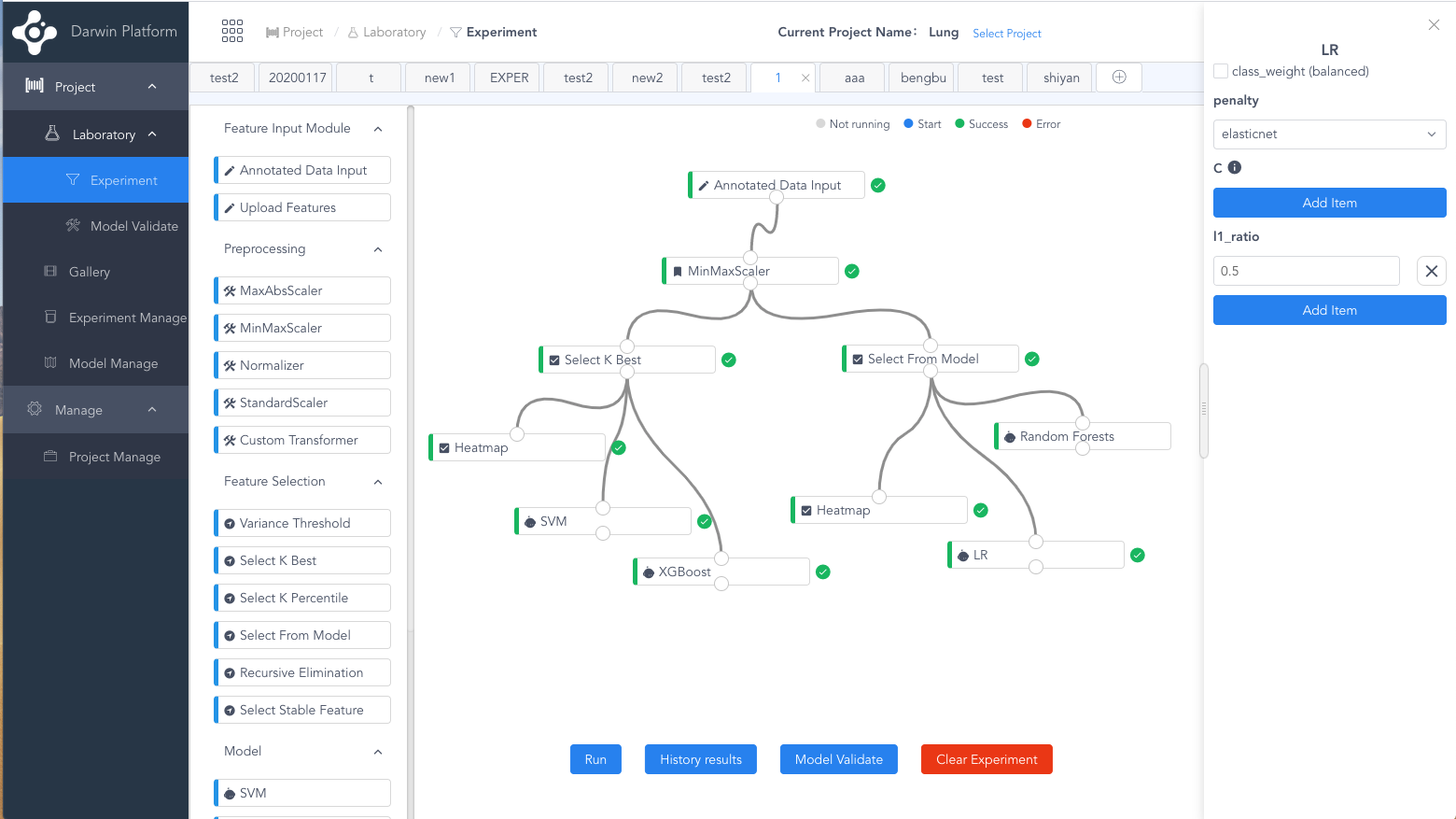}}
\caption{The computational graph of a radiomic experiment. The left sidebar menu presents entries to Laboratory, Image Gallery, and other Projects. The left part of the page is the node-selecting area. Users can drag the nodes to the center and build up a computational graph. All the parameters could be changed on the right side.
}
\label{graph}
\end{figure*}

The main functionalities of DARWIN include the following:
\begin{itemize}

\item{Image Labeler: A GUI for DICOM image loading and display. Users can also draw a region of interest (ROI) and annotate it with a self-defined label (E.g. malignant or benign).
}

\item{Computational Graph: A GUI to build a computational pipeline, which includes image loader, preprocessing, feature extraction, visualization, etc  (Fig \ref{graph}). Deep learning and radiomic experiments have separate computational graphs.
Each computational graph is a tree structure with computational nodes.}

\end{itemize}

\begin{figure}[H]
\centering
\subfigure{
\begin{minipage}[t]{0.3\linewidth}
\centering
\includegraphics[width=0.9\linewidth]{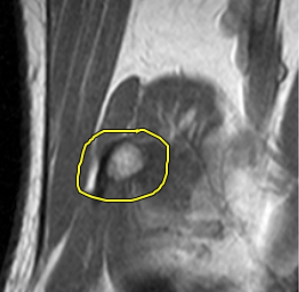}
\end{minipage}}
\subfigure{
\begin{minipage}[t]{0.3\linewidth}
\centering
\includegraphics[width=0.9\linewidth]{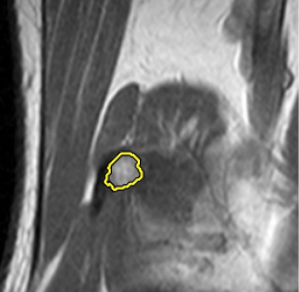}
\end{minipage}}
\subfigure{
\begin{minipage}[t]{0.3\linewidth}
\centering
\includegraphics[width=0.9\linewidth]{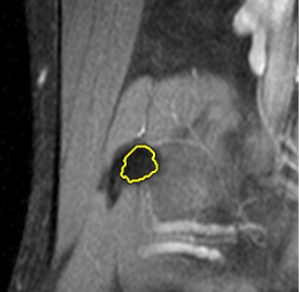}
\end{minipage}}
\caption{Semi-automatic segmentation tools. The first image is a coarse ROI drawn by the user on T1WI. The second image shows the ROI generated by the 'Fit Boundary' tool. The third one is the same ROI automatically transferred to T2WI via the 'Copy/Paste' tool.}
\label{auto-mark}
\end{figure}

\section{Image Labeler and Radiomic Feature Extractor}

The image labeler can load multimodal DICOM images including CT, MRI, CR/DXR, PET, and so on. Like Picture Archiving and Communication System (PACS), we can not only view pictures on it, but also do window level/width adjusting, image scaling, and 3D-reconstruction. ROI is always necessary for analyzing, and we offer tools for ROI loading, drawing, and modifying. For breast cancer and pulmonary nodules, we have built-in convolutional neural networks that can automatically detect and segment lesions. For other types of images and ROIs, the regions can also be drawn manually with the 2D and 3D drawing tool. 
To help physicians draw the ROIs effortlessly, we provided semi-automatic segmentation tools (Fig. \ref{auto-mark}). The 'Fit Boundary' tool implement using the seeded region growing algorithm\cite{adams1994seeded} helps to segment out the bright/dark ROI within a manually drawn curve, and will spread out to 3D neighboring slices if allowed. The 'Copy/Paste' tool allows users to copy an ROI to different series in the same study with automatically registered.

A 2D ROI in our system is a polygon, thus we save it using the position of all its vertexes. Accordingly, it is straightforward for users to modify any vertex. Furthermore, for each ROI, we can annotate labels on it, such as malignant and benign or BI-RADS (0,1,2,3,4,5). An important thing for our labeler is that we can manually or automatically establish some mappings between ROIs.  For example, in some cases, we could have Cranial-Caudal (CC) and mediolateral-oblique (MLO) views for mammograms, and usually, a lesion appears on both views. In our system, We can link the two ROIs to distinguish whether they are from the same lesion, while they share the same annotation information and this relationship should be emphasized in analysis. All of the operations above will be sent to the server and stored in the database for future use. 

\begin{figure}[htbp]
\centerline{\includegraphics[width=0.7\linewidth]{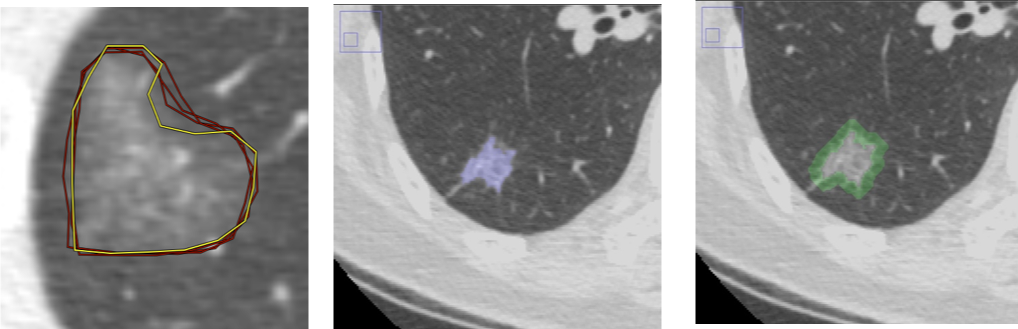}}
\caption{Robust feature extraction. The left one is an example of ROI perturbation. The middle and right ones are visualizations of perilesional features.}
\label{perturb}
\end{figure}

When the user finished labeling one ROI and submitted the results, our system will automatically extract the radiomics features and save them in the database. This operation reduced the delay in radiomic experiments and remarkably improved user experience. We offered more than 1300 dimensions of radiomics features for each ROI. 
The features consist of on average $\sim$ 90 features (Table.\ref{feature_list}) for the original and derived images (8 levels of Wavelet decompositions; Laplacian of Gaussian, Square, Square Root, Logarithm and Exponential filters).
These feature extractors are implemented using the PyRadiomics \cite{pyradiomics} library.
We also provided a robust feature extraction choice (Fig. \ref{perturb}). Users can perform perturbations on ROI and extract perilesional features to get more stable results.

\section{Computational Graph}
A Computational Graph is a directed graph represents an experimental pipeline. The users are able to build a graph using our GUI in several minutes instead of writing scripts in some programming languages, which is non-trivial for radiologists with limited coding experience. Once the graph is established, we can still easily modify it by adding/deleting or replacing some nodes on it. 

\subsection{Graph Nodes}
Each node in the graph is a basic computational unit, it has specified input and output format. A node receives data from one or more nodes and the output will be passed to several nodes. When an error raised on one node, it does not have influences on other independent ones. E.g., the error occurred on leaf nodes will not influence others. 
Deep learning and radiomics laboratories have separate classes of nodes.  

\subsubsection{Radiomics}
\paragraph{Dataloader}
The dataloader class is used to load image features. 
Users can choose to upload the existing features or use the 'Annotated Data Input' node to read the feature information from our database which is generated by the image labeler. The dataloader will also read the annotation and linking information. 
The data loaded will be split into a training set and a validation set randomly. Users can also assign each image for training or validation manually.

\paragraph{Preprocessing}
Data preprocessing is usually needed before feeding data to models. Our platform provides more than five different nodes for data preprocessing. The most common ones are normalizing to $[-1, 1]$ or $[0, 1]$ which are 'Max-Abs Scaler' and 'Min-Max Scaler' respectively. We also present 'Standard Scaler' and 'Normalizer' which represent for projecting all data onto a unit ball and a standard normal distribution separately.
In addition, users are also allowed to write a self-defined transformation function in Python for personal use in the 'Custom Transformer' node.

\paragraph{Feature Selection}
The radiomics feature extractor will output more than 1300 dimensions of features which is usually redundant. Feature selection is an important step that will heavily influence the model performance. To select out the most useful features, we provide numerous feature selection methods. Specifically, we have five different kinds of nodes. (i) Variance Threshold: choosing the features with higher variance. (ii) Select K Best / Select K Percentile: users can set how many features s/he wants and what measure to use. The measure can be Chi-Square statistics, ANOVA-F statistics, or mutual information \cite{lomax2007statistical} between each feature and annotated label. (iii) Select From Model: selecting the features by the weights from a model. We have L1 Lasso, SVM, and many decision tree-based methods. (iv) Recursive Feature Elimination: recursively decrease the feature size according to a model. SVM and other tree-based methods are supported. (v) Select Stable Feature: select k features that stable to different feature selection algorithms.
For any feature selection node, one can single-click the node result to check the features name and corresponding importance score they get from the node when the computation finished. 


\paragraph{Visualization}
Heatmap and dimension reduction are common ways to visualize the overall distribution of high dimension($\ge$ 3) data. They will give a comprehensive impression of the goodness of feature selection results. These dimension reduction methods can also be used to show the results of hierarchical data clustering which is a common unsupervised method for data analyzing. We provided 2 nodes for data visualization, heatmap and T-SNE \cite{tsne}.

\paragraph{Machine Learning Models}
The key component in the computational graph is to choose the best machine learning model to learn the relationship between our processed features and the human annotated labels. We have implemented many classical classifiers including SVM, Logistic Regression, Decision Tree, Gradient Boosting Decision Tree, Random Forest, XGboost, and Adaboost. As for the hyperparameter of each model, we automatically apply grid search cross validation to choose the best ones which will be very convenient for those who are not so familiar with these ML models. After training the model, We will draw the ROC curve and compute the AUC and AP score for the model. Also, the p-values of the hypothesis test for AUC and AP will be reported to show statistical significance. 

\begin{table}[]
    \centering
    \begin{tabular}{cll}
    Radiomic Models   &  Classification & SVM \\
    & & Logistic Regression \\
    & & Decision Tree \\
    & & Random Forest \\
    & & GBDT \\
    & & XGBoost \\
    & Regression & Nomogram\\
    & Clustering & Kmeans\\
    & & HDBSCAN\\
    \hline
    Deep Learning Models  & Classification & VGG-16,19 \\
    & & Resnet-18,34,50,101 \\
    & & Densnet-121,169 \\
    & & Resnext-50, 101 \\
    & & Incpetion V4 \\
    & & Inception-Resnet-V2\\
    & & Xception \\
    & Detection & SSD \\
    & & YOLOv2,v3 \\
    & & Faster RCNN \\
    & & Mask RCNN\\
    & & RetinaNet \\
    & Segmentation& U-net \\
    & & FCN \\
    & & Mask RCNN\\
    \end{tabular}
    \caption{Models Provided in Darwin Platform. Both 2D and 3D images are supported}
    \label{model_list}
\end{table}

\begin{table*}[]
    \centering
    \begin{tabular}{cl|cl}
Shape  & Elongation  &   GLCM & Autocorrelation \\
 & Flatness  &   & ClusterProminence \\
 & LeastAxisLength  &   & ClusterShade \\
 & MajorAxisLength  &   & ClusterTendency \\
 & Maximum2DDiameterColumn  &   & Contrast \\
 & Maximum2DDiameterRow  &   & Correlation \\
 & Maximum2DDiameterSlice  &   & DifferenceAverage \\
 & Maximum3DDiameter  &   & DifferenceEntropy \\
 & MeshVolume  &   & DifferenceVariance \\
 & MinorAxisLength  &   & Id \\
 & Sphericity  &   & Idm \\
 & SurfaceArea  &   & Idmn \\
 & SurfaceVolumeRatio  &   & Idn \\
 & VoxelVolume  &   & Imc1 \\
 Firstorder  & 10Percentile &  & Imc2 \\
 & 90Percentile &  & InverseVariance \\
 & Energy &   & JointAverage \\
 & Entropy &  & JointEnergy \\
 & InterquartileRange &   & JointEntropy \\
 & Kurtosis &  & MCC \\
 & Maximum &   & MaximumProbability \\
 & Mean &   & SumAverage \\
 & MeanAbsoluteDeviation &   & SumEntropy \\
 & Median &  & SumSquares \\
 & Minimum & & \\
 & Range & & \\
 & RobustMeanAbsoluteDeviation & & \\
 & RootMeanSquared & & \\
 & Skewness & & \\
 & TotalEnergy & & \\
 & Uniformity & & \\
 & Variance & & \\
 \hline
 GLDM & DependenceEntropy &   GLRLM  & GrayLevelNonUniformity \\
 & DependenceNonUniformity &   & GrayLevelNonUniformityNormalized \\
 & DependenceNonUniformityNormalized &   & GrayLevelVariance \\
 & DependenceVariance &   & HighGrayLevelRunEmphasis \\
 & GrayLevelNonUniformity &   & LongRunEmphasis \\
 & GrayLevelVariance &   & LongRunHighGrayLevelEmphasis \\
 & HighGrayLevelEmphasis &   & LongRunLowGrayLevelEmphasis \\
 & LargeDependenceEmphasis &   & LowGrayLevelRunEmphasis \\
 & LargeDependenceHighGrayLevelEmphasis &   & RunEntropy \\
 & LargeDependenceLowGrayLevelEmphasis &   & RunLengthNonUniformity \\
 & LowGrayLevelEmphasis &   & RunLengthNonUniformityNormalized \\
 & SmallDependenceEmphasis &   & RunPercentage \\
 & SmallDependenceHighGrayLevelEmphasis &   & RunVariance \\
 & SmallDependenceLowGrayLevelEmphasis &   & ShortRunEmphasis \\
 & &  & ShortRunHighGrayLevelEmphasis \\
 & &  & ShortRunLowGrayLevelEmphasis \\
\hline
 GLSZM  & GrayLevelNonUniformity &  NGTDM  & GrayLevelNonUniformity \\
 & GrayLevelNonUniformityNormalized &  & GrayLevelNonUniformityNormalized \\
 & GrayLevelVariance &  & GrayLevelVariance \\
 & HighGrayLevelZoneEmphasis &  & HighGrayLevelZoneEmphasis \\
 & LargeAreaEmphasis &  & LargeAreaEmphasis \\
 & LargeAreaHighGrayLevelEmphasis &  & LargeAreaHighGrayLevelEmphasis \\
 & LargeAreaLowGrayLevelEmphasis &  & LargeAreaLowGrayLevelEmphasis \\
 & LowGrayLevelZoneEmphasis &  & LowGrayLevelZoneEmphasis \\
 & SizeZoneNonUniformity &  & SizeZoneNonUniformity \\
 & SizeZoneNonUniformityNormalized &  & SizeZoneNonUniformityNormalized \\
 & SmallAreaEmphasis &  & SmallAreaEmphasis \\
 & SmallAreaHighGrayLevelEmphasis &  & SmallAreaHighGrayLevelEmphasis \\
 & SmallAreaLowGrayLevelEmphasis &  & SmallAreaLowGrayLevelEmphasis \\
 & ZoneEntropy &  & ZoneEntropy \\
 & ZonePercentage &  & ZonePercentage \\
 & ZoneVariance  &  & ZoneVariance \\
 \hspace{3pt}
    \end{tabular}
    \caption{Radiomic Features Provided in Darwin Platform}
    \label{feature_list}
\end{table*}

\subsubsection{Deep learning}
\paragraph{Image Input} 
Different from the radiomics module which takes features as input, deep learning module has to take images as input. The Image Input class provides two kinds of nodes, which allows users to choose to use the whole images or only ROI regions for the experiment. Similar to the radiomics module, the training and validation set will be randomly split.

\paragraph{Preprocessing}
Deep learning preprocessing class includes the normalization and standardization node similar to radiomics. It also provides a feature-based image alignment node for multi-modal images. Augmentation is also an essential part of deep learning experiments. We integrated albumentations library \cite{buslaev2020albumentations:} for 2D cases, and implement 3D augmentation functions. More than 20 types of augmentation including Flip, Transpose, Crop, Rotate, Blur, Elastic are supported.

\paragraph{DL models} Our platform support all the 2D and 3D tasks of Detection, Segmentation, and Classification. The classification node provides VGG, Resnet, Densenet, Resnext, Inception, and Xception. The object detection node includes Faster R-CNN, YOLO-v2, SSD, Mask RCNN, and RetinaNet. The segmentation network includes U-net, Mask RCNN, FCN, and its variants. Table \ref{model_list} lists the network architectures currently supported in our platform. More state of the art models will be added in regularly updating.

\paragraph{Training}
In the training class, users can choose to upload the pretrain weights and then train or finetune the network. All the hyper-parameters including the total epochs, batch size, learning rate, learning rate scheduler, optimizer can be manually selected. We also provide hyper-parameter searching algorithms including 
random search, first in first out (FIFO), and hyperband. These algorithms are implemented with Auto-Gluon library \cite{agtabular}. During training, there will be a graph shows information about GPU usage, estimated training time, current iteration steps, training and validation loss, accuracy, AUC scores. This will help users to find potential errors in an early stage. Same as the radiomics module, users are able to view the ROC curve, AUC score, confusion matrix after training is completed.

\paragraph{Visualization}
In deep learning, a commonly used way for visualization is class activation mapping (CAM). In this node, users need to upload the weights, select the target images, and specify the layer to visualize. The generated heatmap will help physicians understand how the network makes a decision, and could carry some potential underlying messages with medical meanings.

\paragraph{Ensemble}
Ensembling is a widely used trick in deep learning competitions. Ensemble a group of models usually gives better performance than any single one. Here we present two kinds of ensemble methods: voting and averaging. Radiomics models are also supported in this ensemble node.

\subsection{Graph Building}
There are three steps to build a graph, selecting the nodes, dragging these nodes to the canvas, and connecting them by mouse. Once the graph is built, the users only need to click the 'Run' button. The graph will be sent to the server and the data flow will pass through each node sequentially until all nodes have finished their computations. After that, one can click each node to check the output of it. 
The connection between all nodes is flexible. For instance, one can feed the radiomics feature to preprocessing nodes, feature selection nodes, and machine learning nodes directly. To compare different feature selection methods simultaneously, s/he could just feed the radiomics features to several feature selection nodes parallelly specified by different functions. The process is similar for machine learning nodes when comparing different models. Additionally, visualization nodes can receive a set of features and corresponding values which means it can follow feature nodes, preprocessing nodes, and feature selection nodes. Fig.~\ref{graph} shows the completed computational graph for a radiomic experiment.

\subsection{Model Retrieval}
Once an experiment completed, the computational graph and corresponding model weights will be saved in the model database. Users can retrieve the history results and compare the performance. For each history record, users are allowed to specify a new test set for the model test.

\section{Experiments}
\subsection{Radiomic Performance Evaluation}
To verify the performance of the radiomic module in the platform, we carried out a demo project of identifying an ROI is on the left or the right lung. We randomly chose 19 images from LIDC-IDRI \cite{armato2011lung} and manually annotated 424 3D ROIs. The experiment runs on a server with 12 cores Intel i7-8700K CPU @ 3.70GHz. The average time for extracting features of one ROI is 1.39 seconds. We use 339 ROIs for training, 85 ROIs for validation. 

\begin{figure}[htbp]
{
\centering{
\includegraphics[width=0.7\linewidth]{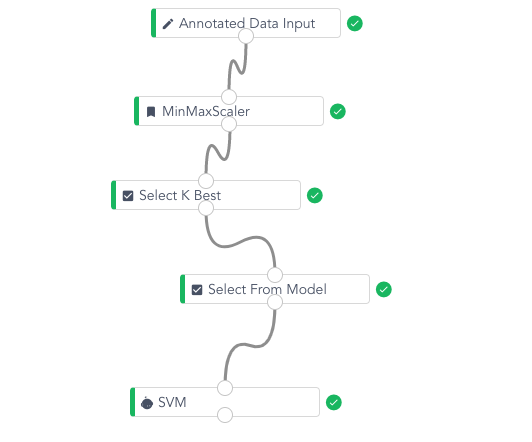} }}
\caption{Computational Graph for Lung ROI Classification.}
\label{lung-graph}
\end{figure}

\begin{figure}[htbp]
{
\centering{
\includegraphics[width=\linewidth]{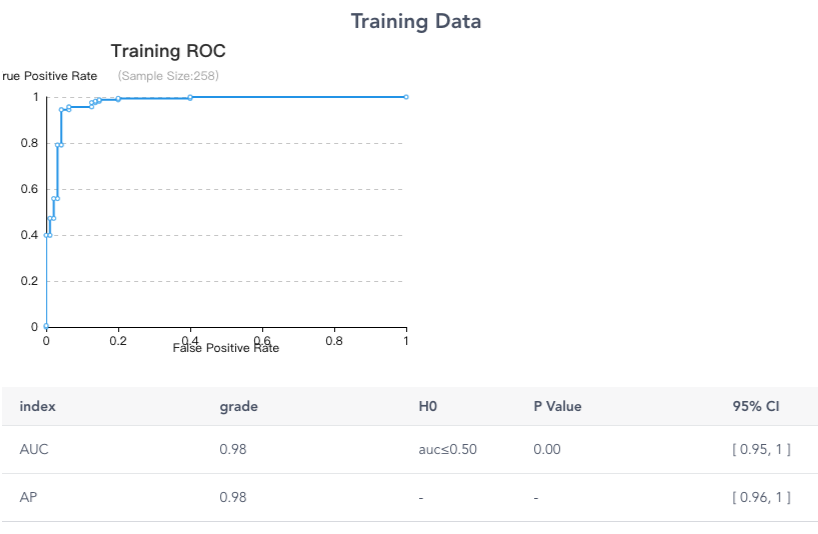} 
\includegraphics[width=\linewidth]{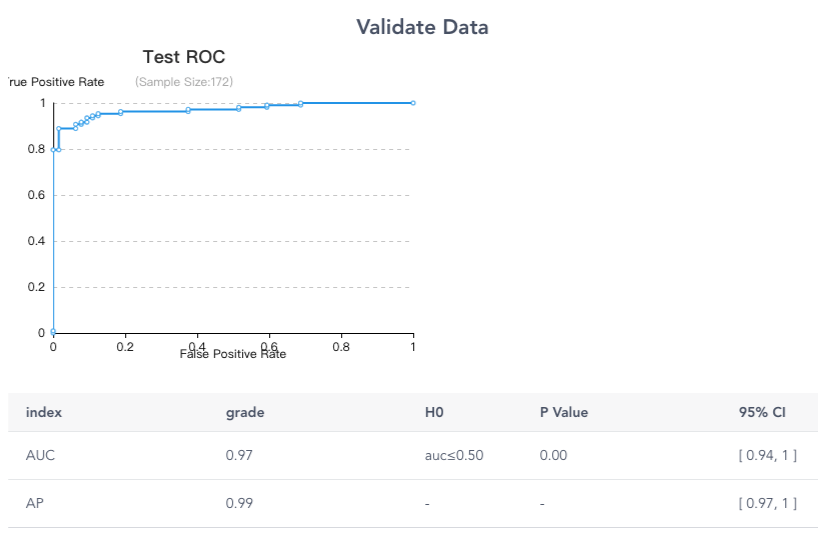} }}
\caption{Lung ROI Classification Results.}
\label{lung-res}
\end{figure}

We constructed a computational graph in Fig.\ref{lung-graph}. Forty features are selected from logistic regression, and then feed to the SVM classifier. The running time of this whole pipeline is 234.25 seconds. Classification results are shown in Fig.\ref{lung-res}. It achieves a high performance with over 0.97 AUC and 0.99 average precision (AP) on the validation set.

\subsection{Deep Learning Performance Evaluation}

\begin{figure}[htbp]
\begin{center}{
\includegraphics[width=0.63\linewidth]{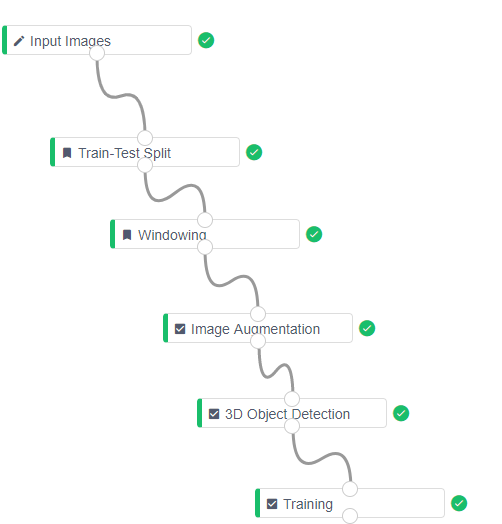}}
\end{center}
\caption{Computational Graph for Lung Nodules Detection.}
\label{deep-graph}
\end{figure} 

We collected 10,701 lung CT scans from 10 hospitals and constructed a model for lung nodules detection. 
The computational graph is shown in Fig. \ref{deep-graph}. 
We randomly chose 260 scans for validation, the rest for training. A 3D RetinaNet model trained from scratch for this experiment. The training pipeline runs 8 hours on a sever with 8 TITAN RTX.

\begin{figure}[htbp]
{
\centering{
\includegraphics[width=0.8\linewidth]{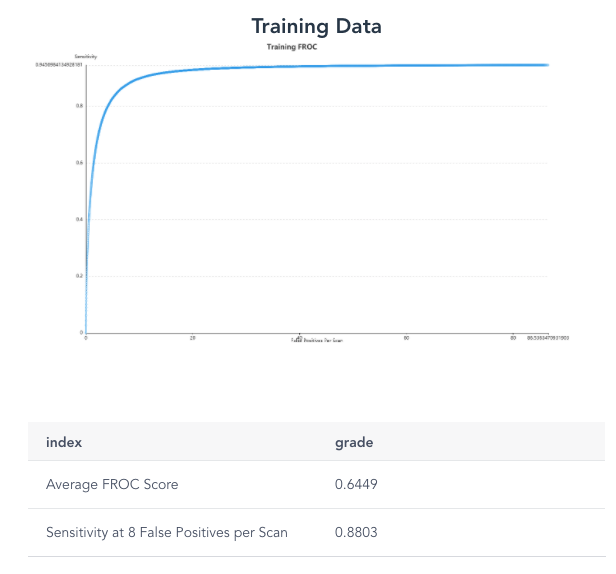} 
\includegraphics[width=0.8\linewidth]{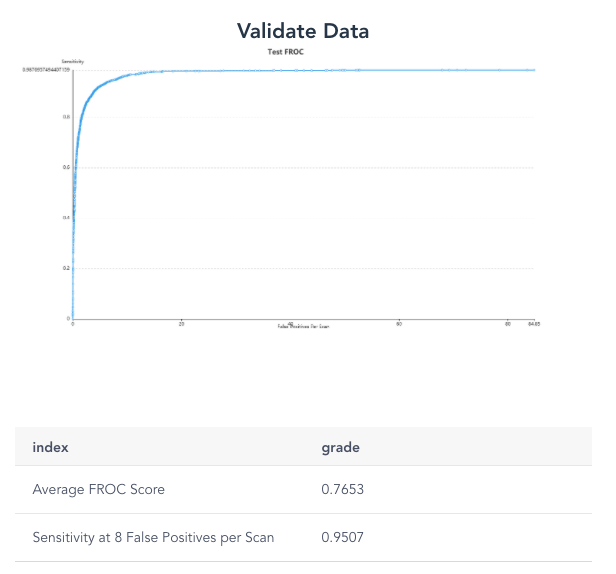} }}
\caption{Lung Nodules Detection Results.}
\label{deep-res}
\end{figure}

Fig.\ref{deep-res} shows the Free-Response ROC (FROC) result. The model achieves a high performance with sensitivity over 0.95 at 8 false positive per scan on the validation set.

\subsection{Case Study: Identifying Recurrent Glioma from Radiation-induced Temporal Lobe Necrosis in Contrast MRI}
Doctors collected examinations of 83 patients for this study. All the patients have tumor-shape areas in their CE-T1WI after 2 weeks of operation.
We balanced split 70\% of the examinations for training, the rest for test. 

\begin{figure}[htbp]
\begin{center}
{
\centering{
\includegraphics[width=0.7\linewidth]{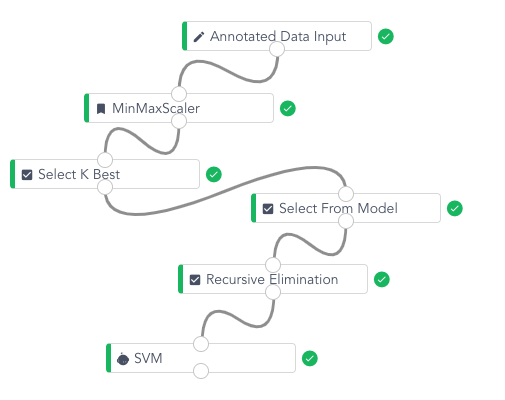} }}    
\end{center}
\caption{The Computational Graph for Glioma and Necrosis Classification.}
\label{glioma-graph}
\end{figure}

The computational graph is shown in Fig.\ref{glioma-graph}. The input 1223 dimension features are reduced to 100 by ANOVA F-value. Doctors further selected 4 features based on the weights of logistic regression and SVM.

\begin{figure}[htbp]{
\centering{
\includegraphics[width=0.9\linewidth]{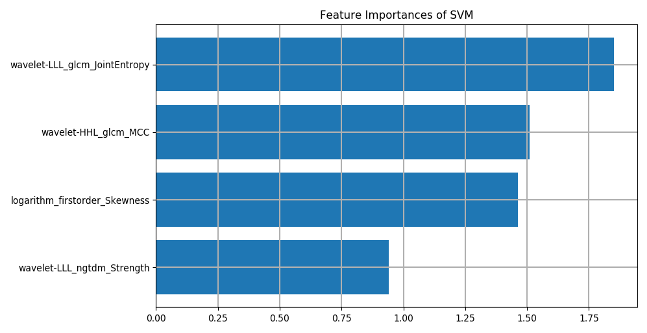} }}
\caption{Features Selected in Glioma and Necrosis Classification.}
\label{glioma-fea}
\end{figure}

 Fig.\ref{glioma-fea} shows the 4 features and the corresponding feature importance in SVM. Tumors usually have a higher level of  heterogeneity than necrosis, thus an ROI with higher entropy is more likely to be a glioma. Maximal correlation coefficient (MCC) is a measure of the complexity of the texture. Skewness describes the brightness distribution of the ROI. Strength is an indication of the image primitives. The 4 features suggest the images of temporal lobe necrosis have a higher level of unbalance in distribution, and have more complex textures. This is consistent with the histological features of the two diseases. When the recurrent glioma occurs, tumor cells grow rapidly and blood vessels are forming \cite{al2010distinguishing}. In radiation injury, multiple factors including telangiectasias, thrombosis, fibrinoid necrosis of vessel walls, hyaline degeneration, and hemorrhage are performing, which led the images with more complex contents \cite{oh2007stereotactic}. The AUC scores for the four features JointEntropy, Strength, MCC, Skewness in glioma classification are 0.78, 0.74, 0.74, 0.72 respectively.

\subsection{Case Study: Predict the Invasibility of Adenocarcinoma in Pure Ground-Glass Nodules. }

In this project, doctors studied the invasibility of adenocarcinoma in pure ground-glass nodules (pGGNs). Doctors collected the CT examinations of 136 patients, uploaded them to our system, and manually segmented ROIs. Our feature extractor automatically extracted 1223 features. 

\begin{figure}[htbp]
{
\centering{
\includegraphics[width=0.7\linewidth]{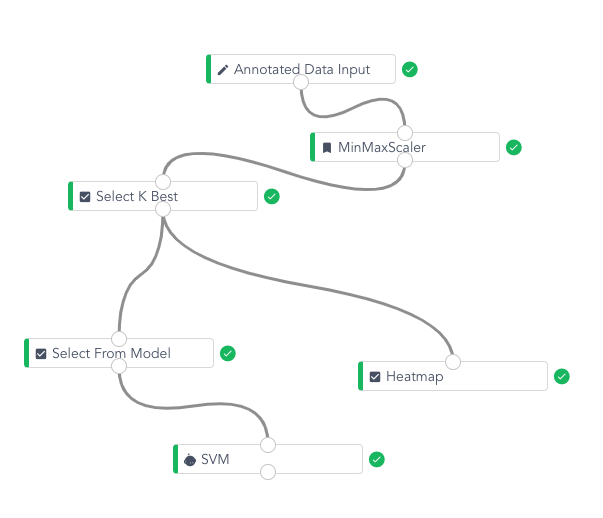} }
}
\caption{The Computational Graph for pGGNs Prediction}
\label{subei-workflow}
\end{figure}

Doctors divided the data into a training set (95 patients), and a test set (41 patients). They constructed a computational graph (Fig. \ref{subei-workflow}) in the laboratory of our Darwin Research Platform.
The 1223 features are normalized into 0-1 for prepossessing. Doctors used Select-K-Best and Recursive-Feature-Elimination using LASSO for feature selection and ultimately got 6 features.
The SVM model is used for classification.

\begin{figure}[htbp]
{
\centering{
\includegraphics[width=0.9\linewidth]{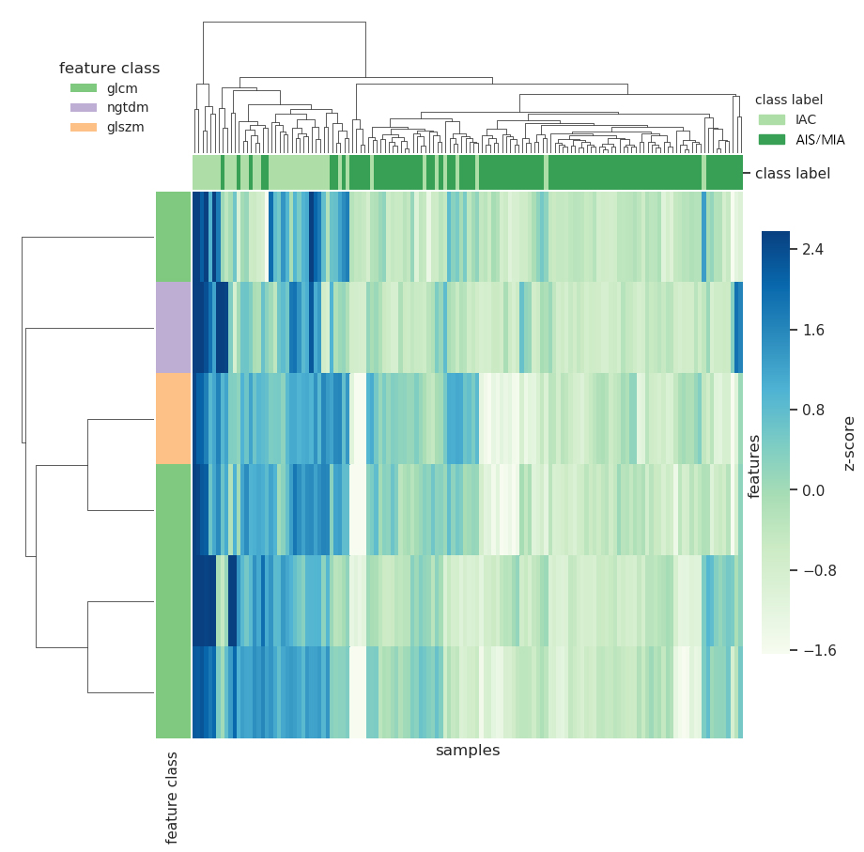} }
}
\caption{Heatmap of the Selected Features in pGGNs prediction}
\label{subei-heatmap}
\end{figure}

\begin{figure}[htbp]{
\centering{
\includegraphics[width=0.8\linewidth]{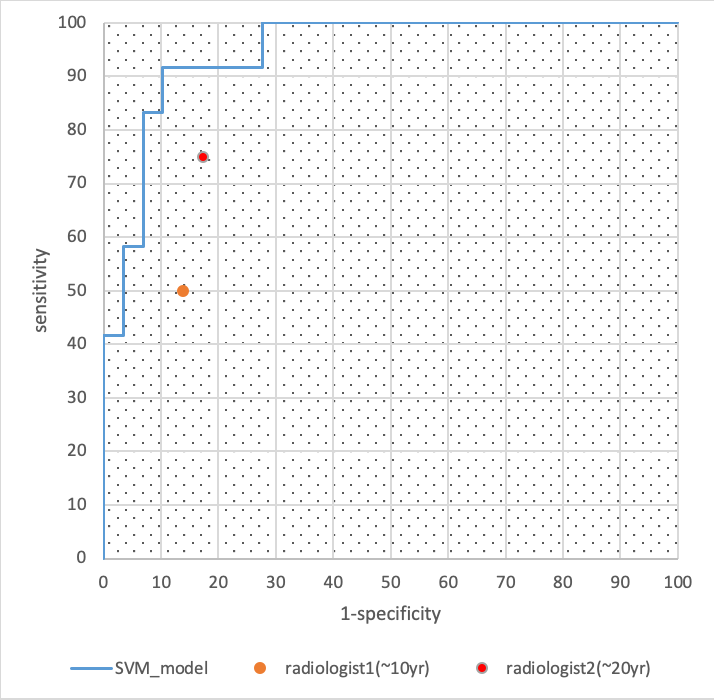} }}
\caption{ROC curve of the SVM model on test set}
\label{subei-roc}
\end{figure}

The platform automatically generated the coefficients and loss paths in the feature selection period.
It also presents the coefficients of the final selected features.
The heatmap of the selected features is shown in Fig. \ref{subei-heatmap}. Table \ref{tab:perform} and Fig. \ref{subei-roc} show the model performance on test set. It achieves an accuracy of over 90\%, which outperform the experienced doctors.

\begin{table}[htbp]
    \centering
    \begin{tabular}{c|c|c|c}
        & Accuracy & Sensitivity & Specificity \\
        \hline
        Radiologist 1 (10-year experience) & 75.61\% & 50.00\% & 86.21\% \\
        Radiologist 2 (20-year experience) & 80.49\% & 75.00\% & 82.76\% \\
        SVM prediction  & 90.24\% & 91.67\% & 89.66\% \\
        
    \end{tabular}
    \caption{Preformance of the SVM model on test set}
    \label{tab:perform}
\end{table}

\section{Conclusion}
In this paper, we present the DARWIN research platform. Radiologists could use the platform to analyze medical imaging data. The DARWIN research platform streamlines the collection, annotation, data preprocessing, model training, model selection, and analysis. The platform provides various choices for users with a user-friendly GUI. Physicians could conduct their experiments just by clicking the mouse and dragging the related icons. The platform provides ways to visualize the high-dimension data, allowing users to comprehend the learning period straightforwardly. The generated experiment reports also facilitate doctors with their research papers, enabling an effortless way for radiologists to carry out machine learning experiments.

\ifCLASSOPTIONcaptionsoff
  \newpage
\fi

\bibliographystyle{IEEEtran}
\bibliography{IEEEfull,mainbib}

\begin{thebibliography}{10}
\providecommand{\url}[1]{#1}
\csname url@samestyle\endcsname
\providecommand{\newblock}{\relax}
\providecommand{\bibinfo}[2]{#2}
\providecommand{\BIBentrySTDinterwordspacing}{\spaceskip=0pt\relax}
\providecommand{\BIBentryALTinterwordstretchfactor}{4}
\providecommand{\BIBentryALTinterwordspacing}{\spaceskip=\fontdimen2\font plus
\BIBentryALTinterwordstretchfactor\fontdimen3\font minus
  \fontdimen4\font\relax}
\providecommand{\BIBforeignlanguage}[2]{{%
\expandafter\ifx\csname l@#1\endcsname\relax
\typeout{** WARNING: IEEEtran.bst: No hyphenation pattern has been}%
\typeout{** loaded for the language `#1'. Using the pattern for}%
\typeout{** the default language instead.}%
\else
\language=\csname l@#1\endcsname
\fi
#2}}
\providecommand{\BIBdecl}{\relax}
\BIBdecl

\bibitem{mckinney2020international}
S.~M. McKinney, M.~Sieniek, V.~Godbole, J.~Godwin, N.~Antropova, H.~Ashrafian,
  T.~Back, M.~Chesus, G.~C. Corrado, A.~Darzi \emph{et~al.}, ``International
  evaluation of an ai system for breast cancer screening,'' \emph{Nature}, vol.
  577, no. 7788, pp. 89--94, 2020.

\bibitem{review}
A.~Vial, D.~Stirling, M.~Field, M.~Ros, C.~Ritz, M.~Carolan, L.~Holloway, and
  A.~A. Miller, ``The role of deep learning and radiomic feature extraction in
  cancer-specific predictive modelling: a review,'' \emph{Translational Cancer
  Research}, vol.~7, no.~3, pp. 803--816, 2018.

\bibitem{wibmer2015haralick}
A.~Wibmer, H.~Hricak, T.~Gondo, K.~Matsumoto, H.~Veeraraghavan, D.~Fehr,
  J.~Zheng, D.~Goldman, C.~Moskowitz, S.~W. Fine \emph{et~al.}, ``Haralick
  texture analysis of prostate mri: utility for differentiating non-cancerous
  prostate from prostate cancer and differentiating prostate cancers with
  different gleason scores,'' \emph{European radiology}, vol.~25, no.~10, pp.
  2840--2850, 2015.

\bibitem{Rapid}
A.~Dekker, S.~Vinod, L.~Holloway, C.~Oberije, A.~George, G.~Goozee, G.~P.
  Delaney, P.~Lambin, and D.~Thwaites, ``Rapid learning in practice: A lung
  cancer survival decision support system in routine patient care data,''
  \emph{Radiotherapy and Oncology}, vol. 113, no.~1, pp. 47--53, 2014.

\bibitem{radiomicML}
C.~Parmar, P.~Grossmann, D.~Rietveld, M.~M. Rietbergen, P.~Lambin, and H.~J.
  Aerts, ``Radiomic machine-learning classifiers for prognostic biomarkers of
  head and neck cancer,'' \emph{Frontiers in oncology}, vol.~5, p. 272, 2015.

\bibitem{Predictout}
S.~H. Hawkins, J.~N. Korecki, Y.~Balagurunathan, Y.~Gu, V.~Kumar, S.~Basu,
  L.~O. Hall, D.~B. Goldgof, R.~A. Gatenby, and R.~J. Gillies, ``Predicting
  outcomes of nonsmall cell lung cancer using ct image features,'' \emph{IEEE
  access}, vol.~2, pp. 1418--1426, 2014.

\bibitem{MLquant}
C.~Parmar, P.~Grossmann, J.~Bussink, P.~Lambin, and H.~J. Aerts, ``Machine
  learning methods for quantitative radiomic biomarkers,'' \emph{Scientific
  reports}, vol.~5, p. 13087, 2015.

\bibitem{shen2015multi}
W.~Shen, M.~Zhou, F.~Yang, C.~Yang, and J.~Tian, ``Multi-scale convolutional
  neural networks for lung nodule classification,'' in \emph{International
  Conference on Information Processing in Medical Imaging}.\hskip 1em plus
  0.5em minus 0.4em\relax Springer, 2015, pp. 588--599.

\bibitem{hussein2017risk}
S.~Hussein, K.~Cao, Q.~Song, and U.~Bagci, ``Risk stratification of lung
  nodules using 3d cnn-based multi-task learning,'' in \emph{International
  conference on information processing in medical imaging}.\hskip 1em plus
  0.5em minus 0.4em\relax Springer, 2017, pp. 249--260.

\bibitem{yan20183d}
K.~Yan, M.~Bagheri, and R.~M. Summers, ``3d context enhanced region-based
  convolutional neural network for end-to-end lesion detection,'' in
  \emph{International Conference on Medical Image Computing and
  Computer-Assisted Intervention}.\hskip 1em plus 0.5em minus 0.4em\relax
  Springer, 2018, pp. 511--519.

\bibitem{kamnitsas2017efficient}
K.~Kamnitsas, C.~Ledig, V.~F. Newcombe, J.~P. Simpson, A.~D. Kane, D.~K. Menon,
  D.~Rueckert, and B.~Glocker, ``Efficient multi-scale 3d cnn with fully
  connected crf for accurate brain lesion segmentation,'' \emph{Medical image
  analysis}, vol.~36, pp. 61--78, 2017.

\bibitem{christ2016automatic}
P.~F. Christ, M.~E.~A. Elshaer, F.~Ettlinger, S.~Tatavarty, M.~Bickel,
  P.~Bilic, M.~Rempfler, M.~Armbruster, F.~Hofmann, M.~D’Anastasi
  \emph{et~al.}, ``Automatic liver and lesion segmentation in ct using cascaded
  fully convolutional neural networks and 3d conditional random fields,'' in
  \emph{International Conference on Medical Image Computing and
  Computer-Assisted Intervention}.\hskip 1em plus 0.5em minus 0.4em\relax
  Springer, 2016, pp. 415--423.

\bibitem{balakrishnan2018unsupervised}
G.~Balakrishnan, A.~Zhao, M.~R. Sabuncu, J.~Guttag, and A.~V. Dalca, ``An
  unsupervised learning model for deformable medical image registration,'' in
  \emph{Proceedings of the IEEE conference on computer vision and pattern
  recognition}, 2018, pp. 9252--9260.

\bibitem{cid2017quantimage}
Y.~D. Cid, J.~Castelli, R.~Schaer, N.~Scher, A.~Pomoni, J.~O. Prior, and
  A.~Depeursinge, ``Quantimage: an online tool for high-throughput 3d radiomics
  feature extraction in pet-ct,'' in \emph{Biomedical Texture Analysis}.\hskip
  1em plus 0.5em minus 0.4em\relax Elsevier, 2017, pp. 349--377.

\bibitem{zhang2015ibex}
L.~Zhang, D.~Fried, X.~Fave, L.~Hunter, J.~Yang, and L.~E. Court, ``Ibex: An
  open infrastructure software platform to facilitate collaborative work in
  radiomics,'' \emph{Medical Physics}, vol.~42, no.~3, pp. 1341--1353, 2015.

\bibitem{nioche2018lifex}
C.~Nioche, F.~Orlhac, S.~Boughdad, S.~Reuze, J.~Goyaouti, C.~Robert,
  C.~Pellotbarakat, M.~Soussan, F.~Frouin, and I.~Buvat, ``Lifex: A freeware
  for radiomic feature calculation in multimodality imaging to accelerate
  advances in the characterization of tumor heterogeneity,'' \emph{Cancer
  Research}, vol.~78, no.~16, pp. 4786--4789, 2018.

\bibitem{yuan2019radiomics}
R.~Yuan, S.~Shi, J.~Chen, and G.~Cheng, ``Radiomics in rayplus: a web-based
  tool for texture analysis in medical images,'' \emph{Journal of Digital
  Imaging}, vol.~32, no.~2, pp. 269--275, 2019.

\bibitem{pyradiomics}
J.~J. Van~Griethuysen, A.~Fedorov, C.~Parmar, A.~Hosny, N.~Aucoin, V.~Narayan,
  R.~G. Beets-Tan, J.-C. Fillion-Robin, S.~Pieper, and H.~J. Aerts,
  ``Computational radiomics system to decode the radiographic phenotype,''
  \emph{Cancer research}, vol.~77, no.~21, pp. e104--e107, 2017.

\bibitem{adams1994seeded}
R.~Adams and L.~Bischof, ``Seeded region growing,'' \emph{IEEE Transactions on
  pattern analysis and machine intelligence}, vol.~16, no.~6, pp. 641--647,
  1994.

\bibitem{lomax2007statistical}
R.~G. Lomax, \emph{Statistical concepts: A second course}.\hskip 1em plus 0.5em
  minus 0.4em\relax Lawrence Erlbaum Associates Publishers, 2007.

\bibitem{tsne}
L.~v.~d. Maaten and G.~Hinton, ``Visualizing data using t-sne,'' \emph{Journal
  of machine learning research}, vol.~9, no. Nov, pp. 2579--2605, 2008.

\bibitem{buslaev2020albumentations:}
A.~V. Buslaev, A.~Parinov, E.~Khvedchenya, V.~Iglovikov, and A.~A. Kalinin,
  ``Albumentations: Fast and flexible image augmentations,''
  \emph{Information-an International Interdisciplinary Journal}, vol.~11,
  no.~2, p. 125, 2020.

\bibitem{agtabular}
N.~Erickson, J.~Mueller, A.~Shirkov, H.~Zhang, P.~Larroy, M.~Li, and A.~Smola,
  ``Autogluon-tabular: Robust and accurate automl for structured data,''
  \emph{arXiv preprint arXiv:2003.06505}, 2020.

\bibitem{armato2011lung}
S.~G. Armato~III, G.~McLennan, L.~Bidaut, M.~F. McNitt-Gray, C.~R. Meyer, A.~P.
  Reeves, B.~Zhao, D.~R. Aberle, C.~I. Henschke, E.~A. Hoffman \emph{et~al.},
  ``The lung image database consortium (lidc) and image database resource
  initiative (idri): a completed reference database of lung nodules on ct
  scans,'' \emph{Medical physics}, vol.~38, no.~2, pp. 915--931, 2011.

\bibitem{al2010distinguishing}
A.~Al~Sayyari, R.~Buckley, C.~McHenery, K.~Pannek, A.~Coulthard, and S.~Rose,
  ``Distinguishing recurrent primary brain tumor from radiation injury: a
  preliminary study using a susceptibility-weighted mr imaging- guided apparent
  diffusion coefficient analysis strategy,'' \emph{American journal of
  neuroradiology}, vol.~31, no.~6, pp. 1049--1054, 2010.

\bibitem{oh2007stereotactic}
B.~C. Oh, P.~G. Pagnini, M.~Y. Wang, C.~Y. Liu, P.~E. Kim, C.~Yu, and M.~L.
  Apuzzo, ``Stereotactic radiosurgery: Adjacent tissue injury and response
  afterhigh-dose single fraction radiation: Part i—histology, imaging, and
  molecularevents,'' \emph{Neurosurgery}, vol.~60, no.~1, pp. 31--45, 2007.

\end{thebibliography}

\end{document}